\documentclass[preprintnumbers,amsmath,amssymbm,prd]{revtex4}
\usepackage{epsfig}
\usepackage{graphicx}
\usepackage{amssymb}

\begin{document}
\title{Charged reflecting stars supporting charged massive scalar field configurations}
\author{Shahar Hod}
\affiliation{The Ruppin Academic Center, Emeq Hefer 40250, Israel}
\affiliation{ }
\affiliation{The Hadassah Institute, Jerusalem 91010, Israel}
\date{\today}

\begin{abstract}
\ \ \ The recently published no-hair theorems of Hod, Bhattacharjee,
and Sarkar have revealed the intriguing fact that horizonless
compact reflecting stars {\it cannot} support spatially regular
configurations made of scalar, vector and tensor fields. In the
present paper we explicitly prove that the interesting no-hair
behavior observed in these studies is not a generic feature of
compact reflecting stars. In particular, we shall prove that charged
reflecting stars {\it can} support {\it charged} massive scalar
field configurations in their exterior spacetime regions. To this
end, we solve analytically the characteristic Klein-Gordon wave
equation for a linearized charged scalar field of mass $\mu$, charge
coupling constant $q$, and spherical harmonic index $l$ in the
background of a spherically symmetric compact reflecting star of
mass $M$, electric charge $Q$, and radius $R_{\text{s}}\gg M,Q$.
Interestingly, it is proved that the discrete set
$\{R_{\text{s}}(M,Q,\mu,q,l;n)\}^{n=\infty}_{n=1}$ of star radii
that can support the charged massive scalar field configurations is
determined by the characteristic zeroes of the confluent
hypergeometric function. Following this simple observation, we
derive a remarkably compact analytical formula for the discrete
spectrum of star radii in the intermediate regime $M\ll
R_{\text{s}}\ll 1/\mu$. The analytically derived resonance spectrum
is confirmed by direct numerical computations.
\end{abstract}
\bigskip
\maketitle

\section{Introduction}

Classical black-hole spacetimes are characterized by one-way
membranes (event horizons) that irreversibly absorb matter and
radiation fields. Event horizons are therefore characterized by
purely ingoing boundary conditions, a remarkable physical property
which naturally suggests, as nicely summarized by Wheeler's no-hair
conjecture \cite{Whee,Car,Bektod}, that asymptotically flat black
holes cannot support spatially regular static field configurations
in their exterior spacetime regions.

The no-hair conjecture for classical black-hole spacetimes has
attracted much attention from physicists and mathematicians over the
years. In particular, the elegant no-hair theorems presented in
\cite{Chas,BekMay,Hart,BekVec} have explicitly proved that
asymptotically flat black holes {\it cannot} support physically
acceptable (spatially regular) static configurations made of scalar,
spinor, or vector fields \cite{Notemsm,Hodrc,Herkr}.

The characteristic absorbing property of the black-hole horizon has
played a key role in the classical no-hair theorems
\cite{Chas,BekMay,Hart,BekVec}. It is therefore quite remarkable
that a recently published theorem has extended the no-scalar hair
property to the physically opposite regime of horizonless compact
stars with {\it reflecting} (rather than {\it absorbing}) boundary
conditions \cite{Hodrec,Hodnww}. These reflecting stars describe
gravitating physical objects whose compact surfaces are
characterized by reflecting Dirichlet ($\Psi=0$ at the compact
surface) or Newmann ($d\Psi/dr=0$ at the compact surface) boundary
conditions [see Eq. (\ref{Eq8}) below]. Moreover, in a very
interesting paper, Bhattacharjee and Sarkar \cite{Bha} have recently
extended the regime of validity of the no-hair theorem for
horizonless spacetimes and proved that compact reflecting stars
cannot support spatially regular configurations made of vector
(spin-1) and tensor (spin-2) fields.

In the present paper we shall explicitly prove that the intriguing
no-hair behavior observed in \cite{Hodrec,Hodnww,Bha} is {\it not} a
generic property of horizonless reflecting stars \cite{Noterfbh}. In
particular, we shall use analytical techniques in order to show that
spherically symmetric asymptotically flat charged stars with
reflecting surfaces {\it can} support spatially regular static
matter configurations made of linearized {\it charged} massive
scalar fields.

Interestingly, below we shall prove that, for a spherically
symmetric compact reflecting star of mass $M$ and electric charge
$Q$, there exists a {\it discrete} set
$\{R_{\text{s}}(M,Q,\mu,q,l;n)\}^{n=\infty}_{n=1}$ of star radii
that can support an external spatially regular charged massive
scalar field of proper mass $\mu$, charge coupling constant $q$, and
spherical harmonic index $l$ \cite{Noteunit,Noteqm}. In particular,
we shall explicitly show that the physical properties of the
composed
charged-reflecting-star-linearized-charged-massive-scalar-field
configurations can be studied {\it analytically} in the intermediate
radii regime $M\ll R_{\text{s}}\ll 1/\mu$.

\section{Description of the system}

We shall analyze the physical and mathematical properties of an
asymptotically flat spacetime which is composed of a central
spherically symmetric reflecting star of radius $R_{\text{s}}$, mass
$M$, and electric charge $Q$, which is linearly coupled to a static
scalar field $\Psi$ of proper mass $\mu$ and charge coupling
constant $q$. The external spacetime of the charged star is
described by the curved spherically symmetric line element
\cite{Chan}
\begin{equation}\label{Eq1}
ds^2=-f(r)dt^2+{1\over{f(r)}}dr^2+r^2(d\theta^2+\sin^2\theta
d\phi^2)\ \ \ \ \text{for}\ \ \ \ r\geq R_{\text{s}}\  ,
\end{equation}
where
\begin{equation}\label{Eq2}
f(r)=1-{{2M}\over{r}}+{{Q^2}\over{r^2}}\  .
\end{equation}

Substituting the mathematical decomposition \cite{Noteom}
\begin{equation}\label{Eq3}
\Psi(r,\theta,\phi)=\sum_{lm}e^{im\phi}S_{lm}(\theta)R_{lm}(r)\
\end{equation}
for the static charged massive scalar field into the characteristic
Klein-Gordon wave equation
\cite{HodPirpam,Stro,HodCQG2,Hodch1,Hodch2}
\begin{equation}\label{Eq4}
[(\nabla^\nu-iqA^\nu)(\nabla_{\nu}-iqA_{\nu})-\mu^2]\Psi=0\
\end{equation}
(here $A_{\nu}=-\delta_{\nu}^{0}{Q/r}$ is the electromagnetic
potential of the spherically symmetric charged star), and using the
metric components of the curved line element (\ref{Eq1}), one finds
the ordinary differential equation
\cite{HodPirpam,Stro,HodCQG2,Hodch1,Hodch2,Notel,Heun,Abram}
\begin{equation}\label{Eq5}
{{d}
\over{dr}}\Big[r^2f(r){{dR_{lm}}\over{dr}}\Big]+\Big[{{(qQ)^2}\over{f(r)}}-(\mu
r)^2-l(l+1)\Big]R_{lm}=0\
\end{equation}
for the radial part of the scalar eigenfunction. A simple inspection
of the radial equation (\ref{Eq5}) for the charged massive scalar
field in the curved spacetime of the spherically symmetric charged
star reveals the fact that it is invariant under the symmetry
transformation $qQ\to -qQ$. We shall henceforth assume without loss
of generality the dimensionless relation
\begin{equation}\label{Eq6}
qQ>0\  .
\end{equation}

The spatially regular charged massive scalar field configurations
that we shall analyze in the present paper are characterized by
normalizable eigenfunctions that decay exponentially at radial
infinity:
\begin{equation}\label{Eq7}
\Psi(r\to\infty)\sim r^{-1-\alpha}e^{-\mu r}\  ,
\end{equation}
where $\alpha\equiv M\mu$ is the dimensionless star-field mass
parameter [see Eq. (\ref{Eq15}) below]. In addition, we shall assume
that the central compact star is characterized by a reflecting
surface of radius $R_{\text{s}}$. In particular, below we shall
consider two types,
\begin{equation}\label{Eq8}
\begin{cases}
\Psi(r=R_{\text{s}})=0 &\ \ \ \ \text{Dirichlet B. C.}\ ; \\
d\Psi(r=R_{\text{s}})/dr=0 &\ \ \ \ \text{Neumann B. C.}\  ,
\end{cases}
\end{equation}
of inner reflecting boundary conditions for the scalar field in the
curved spacetime of the star.

The radial scalar equation (\ref{Eq5}), supplemented by the boundary
conditions (\ref{Eq7}) and (\ref{Eq8}), determines two {\it
discrete} spectra of radii,
$\{R^{\text{Dirichlet}}_{\text{s}}(M,Q,\mu,q,l;n)\}^{n=\infty}_{n=1}$
and
$\{R^{\text{Neumann}}_{\text{s}}(M,Q,\mu,q,l;n)\}^{n=\infty}_{n=1}$,
which, for a given set $\{M,Q,\mu,q,l\}$ of the star-field physical
parameters, characterize the charged reflecting stars that can
support the spatially regular static configurations of the charged
massive scalar fields. Interestingly, in the next section we shall
explicitly prove that, for charged reflecting stars in the
large-radii regime
\begin{equation}\label{Eq9}
R_{\text{s}}\gg M,Q\  ,
\end{equation}
the ordinary differential equation (\ref{Eq5}) for the charged
massive scalar fields in the background of the spherically symmetric
charged star is amenable to an {\it analytical} treatment.

\section{The resonance conditions of the composed charged-reflecting-star-charged-massive-scalar-field
configurations}

In the present section we shall use analytical techniques, which are
valid in the large-radius regime (\ref{Eq9}), in order to derive the
resonance conditions for the discrete sets of radii,
$\{R^{\text{Dirichlet}}_{\text{s}}(M,Q,\mu,q,l;n)\}^{n=\infty}_{n=1}$
and
$\{R^{\text{Neumann}}_{\text{s}}(M,Q,\mu,q,l;n)\}^{n=\infty}_{n=1}$,
which characterize the composed
charged-reflecting-star-linearized-charged-massive-scalar-field
configurations.

It is convenient to define the new radial eigenfunction
\begin{equation}\label{Eq10}
\psi_{lm}=rf^{1/2}(r)R_{lm}\  .
\end{equation}
Substituting (\ref{Eq10}) into (\ref{Eq5}) and neglecting terms of
order $O(M/r^3,Q^2/r^4)$ [see (\ref{Eq9})], one obtains the
Schr\"odinger-like ordinary differential equation \cite{Noteomt}
\begin{equation}\label{Eq11}
{{d^2\psi}\over{dr^2}}+\Big[-\mu^2-{{2M\mu^2}\over{r}}+{{(qQ)^2-(4M^2-Q^2)\mu^2-l(l+1)}\over{r^2}}\Big]\psi=0\
.
\end{equation}
Next, defining the dimensionless radial coordinate
\begin{equation}\label{Eq12}
x=2\mu r\  ,
\end{equation}
one can bring (\ref{Eq11}) into the familiar form
\begin{equation}\label{Eq13}
{{d^2\psi}\over{dx^2}}+\Big[-{1\over4}-{{M\mu}\over{x}}+{{(qQ)^2-(4M^2-Q^2)\mu^2-l(l+1)}
\over{x^2}}\Big]\psi=0\
\end{equation}
of the Whittaker radial differential equation (see Eq. 13.1.31 of
\cite{Abram}).

The general mathematical solution of the radial differential
equation (\ref{Eq13}) is given by (see Eqs. 13.1.32 and 13.1.33 of
\cite{Abram}):
\begin{equation}\label{Eq14}
\psi(x)=e^{-{1\over2}x}x^{{1\over2}+i\beta}\big[A\cdot
U({1\over2}+i\beta+\alpha,1+2i\beta,x)+B\cdot
M({1\over2}+i\beta+\alpha,1+2i\beta,x)\big]\ ,
\end{equation}
where $\{A,B\}$ are normalization constants. Here $M(a,b,z)$ and
$U(a,b,z)$ are respectively the confluent hypergeometric functions
of the first and second kinds \cite{Abram}, and
\begin{equation}\label{Eq15}
\alpha\equiv M\mu\ \ \ \ ; \ \ \ \ \beta^2\equiv
(qQ)^2-(4M^2-Q^2)\mu^2-(l+{1\over2})^2\
\end{equation}
are dimensionless physical parameters which characterize the
composed charged-star-charged-field system. We shall henceforth
assume that \cite{Notedp}
\begin{equation}\label{Eq16}
\beta\in \mathbb{R}\  .
\end{equation}

Using Eqs. 13.1.4 and 13.1.8 of \cite{Abram}, one finds that the
spatial behavior of the radial scalar eigenfunction (\ref{Eq14}) at
radial infinity is given by
\begin{equation}\label{Eq17}
\psi(x\to\infty)=A\cdot x^{-\alpha}e^{-{1\over2}x}+B\cdot
{{\Gamma(1+2i\beta)}\over{\Gamma({1\over2}+i\beta+\alpha)}}x^{\alpha}e^{{1\over2}x}\
.
\end{equation}
Taking cognizance of the asymptotic boundary condition (\ref{Eq7}),
which characterizes the physically acceptable (normalizable)
eigenfunctions of the bound-state massive scalar fields in the
curved spacetime of the central charged star, one concludes that the
normalization constant of the unphysical (exponentially exploding)
term in the asymptotic expression (\ref{Eq17}) should vanish:
\begin{equation}\label{Eq18}
B=0\  .
\end{equation}

We therefore find that the spatially regular bound-state
configurations of the static charged massive scalar fields in the
curved spacetime of the spherically symmetric charged massive star
are characterized by the remarkably compact radial eigenfunction
\begin{equation}\label{Eq19}
\psi(x)=A\cdot e^{-{1\over2}x}x^{{1\over2}+i\beta}
U({1\over2}+i\beta+\alpha,1+2i\beta,x)\ \ \ \ \text{for}\ \ \ \
R_{\text{s}}\gg M,Q\  .
\end{equation}
Taking cognizance of the inner (Dirichlet/Neumann) boundary
conditions (\ref{Eq8}), which characterize the behavior of the
scalar fields at the reflecting compact surface of the spherically
symmetric star, one obtains the two characteristic resonance
equations
\begin{equation}\label{Eq20}
U({1\over2}+i\beta+\alpha,1+2i\beta,2\mu R_{\text{s}})=0\ \ \ \
\text{for}\ \ \ \ \text{Dirichlet B. C.}\
\end{equation}
and
\begin{equation}\label{Eq21}
{{d}\over{dr}}[e^{-\mu r}r^{-{1\over2}+i\beta}f^{-1/2}(r)
U({1\over2}+i\beta+\alpha,1+2i\beta,2\mu r)]_{r=R_{\text{s}}}=0\ \ \
\ \text{for}\ \ \ \ \text{Neumann B. C.}\
\end{equation}
for the composed
charged-reflecting-star-charged-massive-scalar-field configurations
\cite{Notefl,Hodbb}.

Interestingly, as we shall explicitly show in the next section, the
analytically derived resonance conditions (\ref{Eq20}) and
(\ref{Eq21}) determine the characteristic sets of discrete star
radii,
$\{R^{\text{Dirichlet}}_{\text{s}}(M,Q,\mu,q,l;n)\}^{n=\infty}_{n=1}$
and
$\{R^{\text{Neumann}}_{\text{s}}(M,Q,\mu,q,l;n)\}^{n=\infty}_{n=1}$,
which can support the spatially regular bound-state configurations
of the linearized static charged massive scalar fields.

\section{The characteristic resonance spectra of the composed charged-reflecting-star-charged-massive-scalar-field
configurations}

The analytically derived resonance conditions (\ref{Eq20}) and
(\ref{Eq21}), which determine the unique families of charged
reflecting stars that can support the spatially regular static
configurations of the linearized charged massive scalar fields, can
easily be solved numerically. Interestingly, we find that, for a
given set $\{M,Q,\mu,q,l\}$ of the physical parameters that
characterize the composed
charged-reflecting-star-charged-massive-scalar-field system and for
a given reflecting inner boundary condition [see Eq. (\ref{Eq8})],
there exists a {\it discrete} set of star radii,
\begin{equation}\label{Eq22}
\cdots R_{\text{s}}(n=3)<R_{\text{s}}(n=2)<R_{\text{s}}(n=1)\equiv
R_{\text{max}}(M,Q,\mu,q,l)\ ,
\end{equation}
which characterize the compact spherically symmetric charged
reflecting stars that can support the bound-state configurations of
the spatially regular charged massive scalar fields.

From the resonance equations (\ref{Eq20}) and (\ref{Eq21}) one
learns that the dimensionless radii $\{\mu R_{\text{s}}(n)\}$ of the
central supporting stars depend on the dimensionless star-field
physical parameter $\alpha$ and $\beta$ [see Eq. (\ref{Eq15})]. In
Table \ref{Table1} we present, for various values of the
dimensionless star-field mass parameter $\alpha\equiv M\mu$, the
{\it largest} possible dimensionless radii $\{\mu
R_{\text{max}}(\alpha)\}$ of the central charged reflecting stars
that can support the spatially regular configurations of the static
charged massive scalar fields. The data presented in Table
\ref{Table1} reveal the fact that, for a fixed value of the
dimensionless physical parameter $\beta$, the dimensionless star
radii $\{\mu R_{\text{max}}(\alpha)\}$ are a monotonically
decreasing function of the dimensionless star-field mass parameter
$\alpha\equiv M\mu$. We find qualitatively similar results for the
case of reflecting Neumann boundary conditions with the
characteristic property
$R^\text{Neumann}_{\text{max}}(\alpha)>R^\text{Dirichlet}_{\text{max}}(\alpha)$.

\begin{table}[htbp]
\centering
\begin{tabular}{|c|c|c|c|c|c|c|c|}
\hline $\alpha\equiv M\mu\ $ & \ $\ 0.05\ $\ \ & \ $\ 0.1\ $\ \ & \
$\ 0.15\ $\ \ \ & \ $\ 0.20\ $\ \ & \ $\ 0.25\ $\ \ & \ $\ 0.30\ \ $\ \ \\
\hline \ $\ \ \mu R^\text{Dirichlet}_{\text{max}}\ $\ \ \ &\ \
2.3952\ \ \ &\ \ 2.3662\ \ \ &\ \ 2.3377
\ \ \ &\ \ 2.3096\ \ \ &\ \ 2.2819\ \ \ &\ \ 2.2546\ \ \ \\
\hline
\end{tabular}
\caption{Composed
charged-reflecting-star-charged-massive-scalar-field configurations.
We present, for various values of the dimensionless physical
parameter $\alpha\equiv M\mu$, the largest possible dimensionless
radii $\mu R^ \text{Dirichlet}_{\text{max}}(\alpha)$
%and $\mu R^ \text{Neumann}_{\text{max}}(\alpha)$
of the spherically symmetric charged reflecting stars that can
support the static spatially regular configurations of the charged
massive scalar fields. One finds that the maximally allowed radii
$\mu R^ \text{Dirichlet}_{\text{max}}(\alpha)$
%and $\mu R^\text{Neumann}_{\text{max}}(\alpha)$
of the central supporting stars are a monotonically decreasing
function of the star-field dimensionless mass parameter
$\alpha\equiv M\mu$. The data presented is for the case $\beta=5$
[see Eq. (\ref{Eq15})]. Note that the assumption $R_{\text{s}}\gg M$
[see (\ref{Eq9})] is valid for $\alpha\lesssim 0.25$.}
\label{Table1}
\end{table}

In Table \ref{Table2} we present, for various values of the
dimensionless physical parameter $\beta$ [see Eq. (\ref{Eq15})], the
{\it largest} possible dimensionless radii $\{\mu
R_{\text{max}}(\beta)\}$ of the spherically symmetric charged stars
with reflecting Dirichlet boundary conditions that can support the
spatially regular charged massive scalar field configurations. From
the data presented in Table \ref{Table2} one finds that, for a fixed
value of the physical parameter $\alpha$ (which corresponds to a
fixed value of the dimensionless star-field mass parameter $M\mu$),
the dimensionless star radii $\{\mu R_{\text{max}}(\beta)\}$ are a
monotonically increasing function of $\beta$. Again, one finds
qualitatively similar results for the case of reflecting Neumann
boundary conditions with the characteristic property
$R^\text{Neumann}_{\text{max}}(\beta)>R^\text{Dirichlet}_{\text{max}}(\beta)$.

\begin{table}[htbp]
\centering
\begin{tabular}{|c|c|c|c|c|c|c|c|}
\hline $\beta\ $ & \ $\ 1\ $\ \ & \ $\ 3\ $\ \ & \
$\ 5\ $\ \ \ & \ $\ 7\ $\ \ & \ $\ 9\ $\ \ & \ $\ 11\ \ $\ \ \\
\hline \ $\ \ \mu R^\text{Dirichlet}_{\text{max}}\ $\ \ \ &\ \
0.0552\ \ \ &\ \ 0.9803\ \ \ &\ \ 2.3662
\ \ \ &\ \ 3.9205\ \ \ &\ \ 5.5631\ \ \ &\ \ 7.2611\ \ \ \\
\hline
\end{tabular}
\caption{Composed
charged-reflecting-star-charged-massive-scalar-field configurations.
We present, for various values of the dimensionless physical
parameter $\beta$ [see Eq. (\ref{Eq15})], the largest possible
dimensionless radii $\mu R^ \text{Dirichlet}_{\text{max}}(\beta)$
%and $\mu R^ \text{Neumann}_{\text{max}}(\beta)$
of the spherically symmetric charged reflecting stars that can
support the spatially regular configurations of the static charged
massive scalar fields. One finds that the maximally allowed radii
$\mu R^ \text{Dirichlet}_{\text{max}}(\beta)$
%and $\mu R^ \text{Neumann}_{\text{max}}(\beta)$
of the central supporting stars are a monotonically increasing
function of the dimensionless star-field physical parameter $\beta$.
The data presented is for the case $\alpha=0.1$ [$M\mu=0.1$, see Eq.
(\ref{Eq15})], which implies that the assumption $R_{\text{s}}\gg M$
is valid for $\beta\gtrsim 3$.} \label{Table2}
\end{table}

\section{Analytical treatment of the composed star-field system in the intermediate radii regime $M\ll R_{\text{s}}\ll
1/\mu$}

Interestingly, as we shall now prove explicitly, the resonance
equations (\ref{Eq20}) and (\ref{Eq21}), which respectively
determine the discrete sets of radii
$\{R^{\text{Dirichlet}}_{\text{s}}(M,Q,\mu,q,l;n)\}^{n=\infty}_{n=1}$
and
$\{R^{\text{Neumann}}_{\text{s}}(M,Q,\mu,q,l;n)\}^{n=\infty}_{n=1}$
of the charged reflecting stars that can support the spatially
regular bound-state configurations of the charged massive scalar
fields, can be solved {\it analytically} in the intermediate radii
regime
\begin{equation}\label{Eq23}
M\ll R_{\text{s}}\ll 1/\mu\  .
\end{equation}

Using Eqs. 13.1.3 and 13.5.5 of \cite{Abram}, one can express the
(Dirichlet) resonance condition (\ref{Eq20}) of the composed
charged-star-charged-field configurations in the form
%\begin{equation}\label{Eq19}
%{{M({1\over2}+i\beta-\kappa,1+2i\beta,x)}\over{\Gamma({1\over2}-i\beta-\kappa)\Gamma(1+2i\beta)}}=
%x^{-2i\beta}\cdot{{M({1\over2}-i\beta-\kappa,1-2i\beta,x)}\over{\Gamma({1\over2}+i\beta-\kappa)\Gamma(1-2i\beta)}}\
%.
%\end{equation}
\begin{equation}\label{Eq24}
x^{2i\beta}={{\Gamma(1+2i\beta)\Gamma({1\over2}-i\beta+\alpha)}
\over{\Gamma(1-2i\beta)\Gamma({1\over2}+i\beta+\alpha)}}\ \ \ \
\text{for}\ \ \ \ M\ll R_{\text{s}}\ll 1/\mu\  .
\end{equation}
Likewise, using Eqs. 13.1.3, 13.4.21, and 13.5.5 of \cite{Abram},
one can express the (Neumann) resonance condition (\ref{Eq21}) of
the composed charged-star-charged-field configurations in the form
\begin{equation}\label{Eq25}
x^{2i\beta}={{\Gamma(2+2i\beta)\Gamma({1\over2}-i\beta+\alpha)}
\over{\Gamma(2-2i\beta)\Gamma({1\over2}+i\beta+\alpha)}}\ \ \ \
\text{for}\ \ \ \ M\ll R_{\text{s}}\ll 1/\mu\  .
\end{equation}
From the resonance equations (\ref{Eq24}) and (\ref{Eq25}), which
are valid in the intermediate radii regime (\ref{Eq23}), one obtains
the remarkably compact dimensionless resonance spectra
\cite{Notenn,Notesm,Notegr}
\begin{equation}\label{Eq26}
\mu R^{\text{Dirichlet}}_{\text{s}}(n)={{e^{-\pi
n/\beta}}\over{2}}\Big[{{\Gamma(1+2i\beta)\Gamma({1\over
2}-i\beta+\alpha)}\over{\Gamma(1-2i\beta)\Gamma({1\over
2}+i\beta+\alpha)}}\Big]^{1/2i\beta}\ \ \ ; \ \ \ n\in\mathbb{Z}
\end{equation}
and
\begin{equation}\label{Eq27}
\mu R^{\text{Neumann}}_{\text{s}}(n)={{e^{-\pi
n/\beta}}\over{2}}\Big[{{\Gamma(2+2i\beta)\Gamma({1\over
2}-i\beta+\alpha)}\over{\Gamma(2-2i\beta)\Gamma({1\over
2}+i\beta+\alpha)}}\Big]^{1/2i\beta}\ \ \ ; \ \ \ n\in\mathbb{Z}\
\end{equation}
for the {\it discrete} radii of the spherically symmetric charged
reflecting stars that, for a given set of the star-field physical
parameters $\{M,Q,\mu,q,l\}$, can support the spatially regular
bound-state configurations of the static charged massive scalar
fields.

\section{Numerical confirmation}

It is of physical interest to confirm the validity of the
analytically derived discrete resonance spectra (\ref{Eq26}) and
(\ref{Eq27}) which characterize the spherically symmetric charged
reflecting stars that can support the spatially regular external
configurations of the static charged massive scalar fields.

In Table \ref{Table3} we present the dimensionless radii $\{\mu
R^{\text{analytical}}_{\text{s}}(n)\}$ of the spherically symmetric
charged stars with reflecting Dirichlet boundary conditions as
obtained from the analytically derived resonance spectrum
(\ref{Eq26}) in the intermediate radii regime $M\ll R_{\text{s}}\ll
1/\mu$ [see (\ref{Eq23})]. We also present the corresponding
dimensionless radii $\{\mu R^{\text{numerical}}_{\text{s}}(n)\}$ of
the spherically symmetric supporting stars as obtained numerically
from the Dirichlet resonance condition (\ref{Eq20}). From the data
presented in Table \ref{Table3} one finds a remarkably good
agreement \cite{Noteii} between the exact star radii [as computed
numerically from the characteristic resonance condition
(\ref{Eq20})] and the corresponding approximated radii of the
spherically symmetric compact stars [as calculated directly from the
analytically derived discrete resonance spectrum (\ref{Eq26})].

\begin{table}[htbp]
\centering
\begin{tabular}{|c|c|c|c|c|c|c|c|c|}
\hline \text{Formula} & \ $\mu R(n=-5)$\ \ & \ $\mu R(n=-4)$\ \ & \
$\mu R(n=-3)$\ \ & \ $\mu R(n=-2)$\ \ & \ $\mu R(n=-1)$\ \ & \ $\mu R(n=0)$\ \ & \ $\mu R(n=1)$ \\
\hline
\ {\text{Analytical}}\ [Eq. (\ref{Eq26})]\ \ &\ 2.2597\ \ &\ 1.6505\ \ &\ 1.2055\ \ &\ 0.8805\ \ &\ 0.6431\ \ &\ 0.4697\ \ & \ 0.3431\\
\ {\text{Numerical}}\ [Eq. (\ref{Eq20})]\ \ &\ 2.2899\ \ &\ 1.6621\ \ &\ 1.2101\ \ &\ 0.8823\ \ &\ 0.6438\ \ &\ 0.4700\ \ & \ 0.3432\\
\hline
\end{tabular}
\caption{Composed
charged-reflecting-star-charged-massive-scalar-field configurations
with $\alpha\equiv M\mu=0.01$ \cite{Notesm} and $\beta=10$. We
display the discrete resonance spectrum $\{\mu
R^{\text{analytical}}_{\text{s}}(n)\}$ of the dimensionless radii
which characterize the spherically symmetric charged stars with
reflecting Dirichlet boundary conditions as obtained from the
analytically derived resonance formula (\ref{Eq26}) in the
intermediate radii regime $M\ll R_{\text{s}}\ll 1/\mu$. We also
display the corresponding dimensionless radii $\{\mu
R^{\text{numerical}}_{\text{s}}(n)\}$ of the spherically symmetric
supporting stars as obtained numerically from the Dirichlet
resonance condition (\ref{Eq20}). One finds a remarkably good
agreement \cite{Noteii} between the exact star radii [as computed
numerically from the characteristic resonance condition
(\ref{Eq20})] and the corresponding approximated radii of the
spherically symmetric charged reflecting stars [as calculated
directly from the analytically derived compact formula (\ref{Eq26})
for the discrete resonance spectrum].} \label{Table3}
\end{table}

\section{Summary and discussion}

It is well known that asymptotically flat black holes, which are
characterized by compact event horizons with purely ingoing ({\it
absorbing}) boundary conditions, cannot support spatially regular
static configurations made of scalar, spinor, or vector fields
\cite{Whee,Car,Bektod,Chas,BekMay,Hart,BekVec}.

Intriguingly, it has recently been proved that horizonless compact
stars with reflecting (that is, {\it repulsive} rather than {\it
attractive}) boundary conditions cannot support spatially regular
configurations made of neutral scalar (spin-0) fields
\cite{Hodrec,Hodnww}. Moreover, in a very interesting work,
Bhattacharjee and Sarkar \cite{Bha} have later extended the no-hair
theorem of \cite{Hodrec} and proved that horizonless compact stars
with reflective boundary conditions cannot support spatially regular
configurations made of vector (spin-1) and tensor (spin-2) fields.

In the present paper we have proved that the interesting no-hair
behavior revealed in \cite{Hodrec,Hodnww,Bha} for compact reflecting
stars is {\it not} a generic feature of these horizonless objects.
In particular, we have explicitly shown that spherically symmetric
asymptotically flat {\it charged} stars with compact reflecting
surfaces (that is, with Dirichlet/Neumann reflecting boundary
conditions) {\it can} support spatially regular bound-state
configurations made of linearized {\it charged} massive scalar
fields.

Interestingly, we have proved that, for a given set
$\{M,Q,\mu,q,l\}$ of the star-field physical parameters, there exist
two {\it discrete} spectra of radii,
$\{R^{\text{Dirichlet}}_{\text{s}}(M,Q,\mu,q,l;n)\}^{n=\infty}_{n=1}$
and
$\{R^{\text{Neumann}}_{\text{s}}(M,Q,\mu,q,l;n)\}^{n=\infty}_{n=1}$,
which characterize the charged compact reflecting stars that can
support the external bound-state charged massive scalar field
configurations. In particular, it has been explicitly shown that the
physical properties of the composed
charged-reflecting-star-linearized-charged-massive-scalar-field
configurations can be studied {\it analytically} in the regime $M\ll
R_{\text{s}}\ll 1/\mu$ [see Eq. (\ref{Eq23})]. In this intermediate
radii regime we have used analytical techniques in order to derive
the remarkably compact formula [see Eqs. (\ref{Eq26}) and
(\ref{Eq27})]
\begin{equation}\label{Eq28}
\mu R_{\text{s}}={\cal R}\times e^{-\pi n/\beta}\ \ \ ; \ \ \
n\in\mathbb{Z}\
\end{equation}
with
\begin{equation}\label{Eq29}
{\cal
R}(M,Q,\mu,q,l)={1\over2}\Big[\nabla{{\Gamma(1+2i\beta)\Gamma({1\over
2}-i\beta+\alpha)}\over{\Gamma(1-2i\beta)\Gamma({1\over
2}+i\beta+\alpha)}}\Big]^{1/2i\beta}\ \ \ \ ; \ \ \ \
\nabla=\begin{cases}
1 &\ \ \ \ \text{Dirichlet B. C.}\  \\
{{1+2i\beta}\over{1-2i\beta}} &\ \ \ \ \text{Neumann B. C.}\ \
\end{cases}
\end{equation}
for the discrete spectra of radii which characterize the spherically
symmetric charged reflecting stars that can support the spatially
regular charged massive scalar field configurations.

We have further shown that the analytically derived resonance
formula (\ref{Eq28}) for the characteristic discrete spectra of
compact star radii that can support the spatially regular charged
massive scalar field configurations agrees remarkably well
\cite{Noteii} with direct numerical computations of the
corresponding discrete star radii (see Table \ref{Table3}).

Finally, it is important to emphasize again that in the present
study we have treated the spatially regular bound-state scalar
configurations at the linear (perturbative) level. As we explicitly
demonstrated in this paper, the main advantage of this small
amplitude (linearized) treatment stems from the fact that the
physical properties of the composed
charged-reflecting-star-charged-massive-scalar-field system can be
studied {\it analytically} in the linear regime. We believe,
however, that it would be physically interesting to use more
sophisticated numerical techniques in order to explore the physical
properties of these composed star-field configurations in the fully
non-linear regime.

\newpage

\bigskip
\noindent
{\bf ACKNOWLEDGMENTS}
\bigskip

This research is supported by the Carmel Science Foundation. I would
like to thank Yael Oren, Arbel M. Ongo, Ayelet B. Lata, and Alona B.
Tea for helpful discussions.

%\newpage


\begin{thebibliography}{99}
\bibitem{Noteunit} We shall use natural units in which $G=c=\hbar=k_{\text{B}}=1$.
\bibitem{Noteqm} Note that the physical parameters $\mu$ and $q$, which
characterize the charged massive scalar field [see Eq. (\ref{Eq4})
below], stand respectively for $\mu/\hbar$ and $q/\hbar$. These
field parameters therefore have the dimensions of $($length$)^{-1}$.
\bibitem{Noteom} Here the integer parameters $l$ and $m$ (with $-l\leq m\leq l$) are
respectively the spherical harmonic index and the azimuthal harmonic
index of the linearized charged massive scalar field mode.
\bibitem{Notel} Here the integer $l(l+1)$ is the dimensionless eigenvalue of the
spherical harmonic eigenfunction $S_{lm}(\theta)$ which
characterizes the angular behavior of the spatially regular charged
massive scalar field mode \cite{Heun,Abram}.
\bibitem{Noteomt} For brevity, we shall henceforth omit the angular harmonic indices $\{l,m\}$
of the static linearized charged massive scalar field modes.
\bibitem{Notedp} We shall assume, without loss of generality, that $\beta>0$.
\bibitem{Notefl} It is worth noting that, in the dimensionless limit
$\text{max}\{M,Q\}\cdot\mu\to0$ of a flat Minkowski spacetime with
$\alpha=0$ [see Eq. (\ref{Eq15})], one can use the characteristic
mathematical relation
$U(1/2+\nu,1+2\nu,2z)=\pi^{-1/2}e^z(2z)^{-\nu}K_{\nu}(z)$ (see Eq.
13.6.21 of \cite{Abram}) in order to reduce the analytically derived
resonance condition (\ref{Eq20}) to the flat-space resonance
equation $K_{i\beta}(\mu R_{\text{s}})=0$ derived in \cite{Hodbb}
for the static case, where $K_{\nu}(z)$ is the modified Bessel
function of the second kind \cite{Abram}.
\bibitem{Notenn} Here we have used the relation $1=e^{-2i\pi n}$, where the integer
$n$ is the resonance parameter which characterizes the composed
charged-reflecting-star-linearized-charged-massive-scalar-field
configurations.
\bibitem{Notesm} Since each inequality in the intermediate radii regime (\ref{Eq23})
roughly corresponds to an order-of-magnitude difference between two
physical parameters (that is, $M/R_{\text{s}}\lesssim 10^{-1}$ and
$\mu R_{\text{s}}\lesssim 10^{-1}$), the analytically derived
resonance spectra (\ref{Eq26}) and (\ref{Eq27}) for the discrete
star radii are expected to be valid in the small mass regime
$M\mu\lesssim 10^{-2}$.
\bibitem{Notegr} It is worth noting that, using Eq. 6.1.23 of \cite{Abram},
one finds the characteristic dimensionless relations
$\Gamma(1+2i\beta)/\Gamma(1-2i\beta)=e^{i\phi_1}$,
$\Gamma(2+2i\beta)/\Gamma(2-2i\beta)=e^{i\phi_2}$, and
$\Gamma(1/2-i\beta+\alpha)/\Gamma(1/2+i\beta+\alpha)=e^{i\phi_3}$
for $\beta\in \mathbb{R}$ [see Eq. (\ref{Eq16})], where
$\{\phi_1,\phi_2,\phi_3\}\in\mathbb{R}$. These relations imply that
$\{\mu R^{\text{Dirichlet}}_{\text{s}}(n),\mu
R^{\text{Neumann}}_{\text{s}}(n)\}\in\mathbb{R}$ [see Eqs.
(\ref{Eq26}) and (\ref{Eq27})].
\bibitem{Noteii} It is interesting to point out that the data presented in Table
\ref{Table3} reveal the fact that the agreement between the
approximated (analytically derived) resonance formula (\ref{Eq26})
for the discrete spectrum of star radii and the corresponding
numerically computed supporting star radii [as obtained numerically
from the characteristic resonance condition (\ref{Eq20})] is quite
good already in the regime $\mu R_{\text{s}}=O(1)$. This observation
is quite surprising since the analytically derived resonance spectra
(\ref{Eq26}) and (\ref{Eq27}) are formally valid in the small
mass-radii regime $\mu R_{\text{s}}\ll1$ [see (\ref{Eq23})].
\end{thebibliography}
\end{document}